\documentclass[apj,twocolumn]{emulatearxivnew}
\usepackage{amsmath}

\begin{document}

\title{Accretion-Induced Collapse of Dark Matter Admixed White Dwarfs - I: Formation of Low-mass Neutron Stars}

\author{Shing-Chi Leung\thanks{Email address: shingchi.leung@ipmu.jp}} 

\affiliation{Kavli Institute for the Physics and 
Mathematics of the Universe (WPI),The University 
of Tokyo Institutes for Advanced Study, The 
University of Tokyo, Kashiwa, Chiba 277-8583, Japan}

\author{Shuai Zha\thanks{Email address: szha@phy.cuhk.edu.hk}}

\affiliation{Department of Physics, the Chinese
University of Hong Kong, Hong Kong S. A. R., China}

\author{Ming-Chung Chu\thanks{Email address: mcchu@phy.cuhk.edu.hk}}

\affiliation{Department of Physics, the Chinese
University of Hong Kong, Hong Kong S. A. R., China}

\author{Lap-Ming Lin\thanks{Email address: lmlin@phy.cuhk.edu.hk}}

\affiliation{Department of Physics, the Chinese
University of Hong Kong, Hong Kong S. A. R., China}

\author{Ken'ichi Nomoto\thanks{Email address: nomoto@astron.s.u-tokyo.ac.jp}}

\affiliation{Kavli Institute for the Physics and 
Mathematics of the Universe (WPI),The University 
of Tokyo Institutes for Advanced Study, The 
University of Tokyo, Kashiwa, Chiba 277-8583, Japan}

\date{\today}

\begin{abstract}

Recently observed pulsars with masses
$\sim 1.1 ~M_{\odot}$ challenge the conventional neutron star (NS)
formation path by core-collapse supernova (CCSN).
Using spherically symmetric hydrodynamics simulations, we follow the collapse of a massive
white dwarf (WD) core triggered by electron capture, until the formation of a 
proto-NS (PNS). 
For initial WD models with the same central density, 
we study the effects of a static, compact dark matter (DM) admixed core on the collapse and bounce dynamics and 
mass of the PNS, with DM mass $\sim 0.01 ~M_{\odot}$. We show that increasing the admixed DM mass generally leads to slower collapse and smaller PNS mass, down to about 1.0 $M_{\odot}$. 
Our results suggest that the accretion-induced collapse of dark matter admixed white dwarfs can produce
low-mass neutron stars, such as the observed low-mass pulsar J0453+1559, which cannot be obtained by conventional NS formation path by CCSN.

\end{abstract}




\section{Introduction}
\subsection{Dark Matter Physics}

Dark matter (DM) contributes to more than 80 \% 
of mass in the universe \citep{Perlmutter1999, Riess1998, Jarosik2011}. 
The existence of DM is supported by the striking resemblance of the N-body simulation to the observed
large-scale structures \citep{Spergel2005a, Spergel2005b},  
observed galactic rotation curves \citep{Salucci2007},
formation of dwarf galaxies \citep{Baumgardt2008}, and
gravitational weak lensing data \citep{Massey2007}. 

The need for DM in cosmology has its counterpart
in elementary particle physics. The existence of new particles beyond the standard
model is required by a number of theoretical proposals to 
solve open problems in the standard model, including
the strong charge-parity violation problem \citep{Peccei1977},
the need for a mass term for the neutrinos 
and the gauge hierarchy problem \citep{Feng2010}.
They led to proposals of axion, (right-handed) sterile neutrinos
and WIMP (weakly interacting massive particles) respectively. 

Direct hunts for DM particles
such as the DAMA \citep{Bernabei2008}, 
Large Underground Xenon (LUX) \citep{Akerib2013, Akerib2016},
XENON100 \citep{Aprile2010, Aprile2014} and XENON1T \citep{Aprile2019} experiments
so far have not obtained convincing positive results, but the non-detection imposes strong
constraints on DM particle properties \citep{Aprile2016}. 
The DAMA/LIBRA experiment 
has found annual-modulation of the number of detected
events \citep{Bernabei2013, Bernabei2014}, 
but there is no corresponding observation
from the XENON100 and LUX experiments \citep{Aprile2017, Akerib2018}. 

One can also constrain DM properties
by their dark matter's impact on stellar objects. DM
particles admixed into a stellar object in general affect it by providing 
additional gravity or energy sources if they self annihilate or decay.

DM particle self-annihilation can be a prominent energy
source, especially near the galactic center, where the 
DM is concentrated. This can alter 
the evolutionary paths of stellar objects 
\citep{Scott2009, Casanellas2009, Casanellas2011}. 
For compact stars, which have no active nuclear burning, 
the DM self-annihilation energy is a distinctive energy source
and alters their cooling curves and neutrino signals
\citep{Cermeno2018}. In the vicinity of the galactic center,
the DM density is high enough to support DM annihilation
in a WD such that its surface temperature has a distinctive cooling curve
\citep{Moskalenko2007}.
The DM can also be a seed to enhance star
formation and even the first energy source before the
main-sequence (MS) H-burning \citep{Spolyar2009, Freese2008, Freese2009, Hurst2015}, known
as the dark star scenario \citep{Freese2016}. 
For axions, which can convert into photons
in a strong magnetic field, their collisions with a neutron
star (NS) can create a fast radio burst \citep{Iwazaki2015,Raby2016,Clough2018}.

Non-self-annihilating DM \citep{Addazi2015} manifests itself through
its gravity \citep{Goldman1989, Sandin2010}. 
Stars can acquire DM by scattering
through weak interaction \citep{Kouvaris2008}. 
The scattering of DM can lower the energy of 
baryonic matter, causing extra stellar cooling \citep{Zentner2011}.
The energy loss by scattering with DM particles without annihilation may 
resolve the solar composition problem \citep{Frandsen2010}. 
Accretion of DM increases the mass of a stellar object
such as the Sun \citep{Iorio2010} and pulsars \citep{Iorio2010b},
and the orbital paths of its planets would be changed. 
Pure DM can also form self-gravitating objects such as dark matter compact objects
\citep{Narain2006, Kouvaris2015}, axion stars \citep{Barranco2013} and compact planets \citep{Tolos2015}.
Merger events of these compact objects can also produce gravitational
waves just like NS-NS or black hole-black hole mergers
\citep{Bezares2018}. 
However, a recent search of such compact objects through gravitational-wave 
signals has posed a strong limit on their abundances
\citep{Abbott2018}. The gravity of DM admixed in stellar objects, 
such as the Sun \citep{Frandsen2011}, white dwarfs \citep{Leung2013} 
and neutron stars \citep{Leung2011,Leung2012,Ciarcelluti2011,Sandin2010,Rezaei2018},  generally
alter their equilibrium structures. 
This leads to lower Chandrasekhar 
masses for WD and NS, resulting in unusual 
configurations for the subsequent explosion and collapse \citep{Leung2015b,Graham2015, Froggatt2015}.

\subsection{Physics of Accretion Induced Collapse of White Dwarfs}

Accretion-induced collapse (AIC) is the collapse of a WD 
due to accretion of matter from its companion, usually
an MS star. Depending on the accretion rate, the accreted
matter can burn hydrostatically without triggering
nuclear runaways, or the latter are suppressed
by the electron capture that occurs later \citep{Nomoto1982b, Nomoto1991}.
It is a possible path to produce the low-mass 
branch of the bimodal mass distribution of NS discovered in a recent survey \citep{Schwab2010}. 
 
AIC is known to be a quiet event.
The collapse of a WD leads to formation of a NS, 
which gives weak electronmagnetic and gravitational-wave
signals \citep{Metzger2009a,Abdikamalov2010}. 
The AIC of a WD emits
optical, neutrino and gravitational-wave signals and
shows similar behaviour in its 
collapse as the iron-core collapse of a massive 
star or electron capture supernova. 
The shock quickly stalls 
while propagating outwards inside the proto-neutron star (PNS) \citep{Baron1987}. 
Only a small amount of matter can be ejected 
\citep{Canal1976,Mayle1988,Woosley1992}. 
The amount of $^{56}$Ni synthesized during the explosion
is small in the high velocity ejecta \citep{Dharba2010}. 
Such transient objects are believed to be the origins of
some observed gamma-ray bursts \citep{Yi1998} and milli-second
pulsars \citep{Freire2014}. 
In \cite{Abdikamalov2010} two-dimensional
AIC models with general relativistic hydrodynamics but without
neutrino transport
are presented where the gravitational-wave signals are recognized
as the subclass Type III. Similar calculations, but including
neutrino transport, are done in \cite{Dessart2006, Dessart2007} 
for AIC without or with magnetic field respectively. 
 
Another possible channel to produce low-mass
NSs is by SAGB stars (8 - 10 $M_{\odot}$ stars
but the exact mass range is metallicity dependent) 
and their electron capture supernova events
\citep{Nomoto2017b,Leung2019}. The final ONeMg core can have a 
density as high as $10^{9.95}$ g cm$^{-3}$ when 
the nuclear runaway starts \citep{Schwab2015,Nomoto1991,taka13},
where the O-Ne deflagration is triggered by the heating effect
during electron capture of $^{20}$Ne and $^{24}$Mg. 
Unlike Type Ia supernovae, the deflagration
cannot disrupt the star because the high density favours 
electron capture, which impedes the propagation of the deflagration wave. 
In \cite{Leung2017b,Nomoto2017b, Leung2018b} the collapse
scenario is favored for the ONeMg core after the oxygen-neon deflagration.

\subsection{Motivations}

We studied the effects of DM on
stellar evolution and stellar structure extensively in our
previous works \citep{Leung2011, Leung2012, Leung2013,
Leung2015b}. In particular, we have shown that the 
Chandrasekhar mass of compact objects can be significantly
lowered by the presence of  
the DM. A ~0.04 $M_{\odot}$ of DM admixture
can lower the Chandrasekhar mass by 30 $\%$ \citep{Leung2013}.
This makes the corresponding Type Ia supernova explosion 
much weaker than that of the standard Chandrasekhar mass model 
at $\sim 1.4 ~M_{\odot}$. Such weak explosions are further
shown to be consistent with some Type Iax supernovae \citep{Leung2015b}.
The recent discovery of a low-mass binary pulsar system 
J0453+1559 \citep{Martinez2015}
with the companion mass 
as low as $M = 1.174 ~M_{\odot}$
posed a strong challenge to the standard neutron star formation
picture. In view of our previous results, and the possibility
that a WD can undergo AIC to form a NS, 
it becomes interesting to explore the feasibility of 
using this framework to explain the origin of this low-mass NS. 
In fact, as reported in \cite{Leung2013}, the 
Chandrasekhar mass of the progenitor WD can be suppressed down from 1.0 to 0.6 $M_{\odot}$ 
(see Figure 3 in \cite{Leung2013}),
while the corresponding central density increases from 
$\sim 10^9$ to $\sim 10^{11}$ g cm$^{-3}$ (also see Figure 7 in \cite{Leung2013}).
Such a high central density may be the clue for these
low-mass objects to collapse and form low-mass NS.
The current work is a natural extension of our previous
studies on the static properties of DM admixed
compact stars. As far as we are aware, this is also the
first attempt to model the formation of DM admixed NS
proposed by some of us in \cite{Leung2011} using
fully nonlinear hydrodynamical simulations.

In this paper, we first review the numerical code we use 
for the AIC simulation in Section 
\ref{sec:hydro}. Then, in Section \ref{sec:results}
we discuss how the collapse and bounce process and its consequent
PNS are affected by the DM admixture. Unless otherwise noted, we use DM with
a particle mass of 1 GeV in our simulations. We further
examine the robustness of our results to different input physics. 
In Section \ref{sec:discussion} we discuss the applications
of our model to the formation of low-mass NS. We also
discuss what might happen if the DM particle mass is
different 1 GeV.
Finally we present our summary and conclusion.
In the Appendix, we also provide the resolution test to 
check the dependence of our results on the simulation grid size.

\section{Methods}
\label{sec:hydro}

\subsection{Code Update}

We use the one-dimensional version of our hydrodynamics
code, which was originally designed to model SNe Ia. 
We refer the readers to \cite{Leung2015a, Nomoto2017a, Nomoto2017b, Leung2018b} 
for a detailed instrumentation report and the related tests done on 
the code. 

For the initial model, we set up an isothermal WD with 
admixed DM in hydrostatic equilibrium. We solve
\begin{eqnarray}
\frac{dP_{{\rm NM}}}{dr} = -\frac{G m(r) \rho_{{\rm NM}}}{r^2}, \\
\frac{dP_{{\rm DM}}}{dr} = -\frac{G m(r) \rho_{{\rm DM}}}{r^2},
\end{eqnarray}
where \begin{equation}
\frac{dm(r)}{dr} = 4 \pi r^2 (\rho_{{\rm NM}} + \rho_{{\rm DM}}).
\end{equation} 
$\rho_{{\rm NM}} (\rho_{\rm DM})$, $P_{{\rm NM}} (P_{\rm DM})$ are the density and pressure of normal matter (i.e., baryons and leptons) (DM) respectively.
We assume the WD to have a uniform temperature of 0.01 or 0.1 MeV.
For comparison, we also used a more realistic temperature profile,
such as that in \cite{Dessart2006,Abdikamalov2010}:
\begin{equation}
T(\rho_{{\rm NM}}) = T_c (\rho_{{\rm NM}} / \rho_c)^n,
\end{equation}
with $\rho_c$ being the central baryon density, 
$T_c$ = $10^{10}$ K and $n = 0.35$. We refer to the 
model with a non-uniform temperature as the $hot$ model
and the isothermal model to be the $cold$ model. 

To close the Euler equations, we use the SFHo equation of state (EOS) in most of the cases \citep{Steiner2013}. We also use 
the HShen EOS \citep{Shen1998,Shen2011} for comparison.
We choose the SFHo EOS because it can give rise to a 2 $M_{\odot}$ NS, which is 
compatible with PSR J0348+0432 discovered
in \cite{Antoniadis2013}. The compressibility is chosen to be
consistent with constraints from recently observed cooling tracks of X-ray
bursts from NS's in some low-mass X-ray binaries in \cite{Nattila2016}. We also calculate with 
the HShen EOS because of its previous 
applications in AIC \citep{Abdikamalov2010}. 
The Lattimer and Swesty EOS \citep{Lattimer1991} with $K = 220$ MeV (LS220)
is also included for comparison.
For DM, we use the ideal degenerate Fermi gas with a particle mass 1 GeV.

In our modeling, we only 
follow the motion of baryonic matter.
At early time, the DM has a compact core
with a central density much higher than that of the baryonic
matter. The DM core size is much smaller than the whole WD. 
This means that during the collapse of the baryonic matter,
the DM is not affected by the motion of baryonic matter to good approximation.
At later time, when the PNS has formed, 
the baryonic matter has a typical mass density much higher 
than that of DM because of the small amount of DM 
we have used in our models. The motion of the baryonic matter
is then dominated by its own gravity, with DM being a 
small perturbation compared to the baryonic part.  
In both cases, it suffices
to include the DM as a gravity source without explicitly 
evolving its dynamics. Notice that such approximation
may breakdown, if the DM has an initial density or total mass comparable 
to the baryonic matter, or when the total DM mass
is comparable with baryonic matter. 
In these cases, the dynamical timescales for
DM and NM become comparable, and
dynamical modeling of the DM becomes 
important. In our work, there exists
a short period of time during the collapse during which the baryonic matter
has a mass density comparable with that of the DM. But since this occurs in a very  very short duration ($< 10^{-4}$ s), 
we regard our treatment of the DM as a static core a reasonable approximation.  

\subsection{Gravity Solver}

Due to the compactness of the final PNS, 
the previous implementation of Newtonian Gravity \citep{Leung2017b}
will be less accurate. Therefore, 
without changing the code drastically for accommodating the 
metric variables from general relativity, we use the
approximation proposed in \cite{Marek2006}. 
Here we briefly outline the method.

The idea is based on the Tolman - Oppenheimer - Volkoff (TOV)
equation. In the hydrostatic limit, the effective potential can be described
as 
\begin{eqnarray}
\label{eq:TOV}
\Phi_{{\rm TOV}} (r) = -4 \pi \int^{\infty}_r \frac{dr'}{r'^{2}} \left( \frac{m_{{\rm TOV}}}{4 \pi} + r'^{3} (P + p_{\nu}) \right) \times \\
\nonumber \frac{1}{\Gamma^2} \left( \frac{\rho + e + P}{\rho} \right).
\end{eqnarray}
Here, $\rho = \rho_{{\rm NM}} + \rho_{{\rm DM}}$, $e$ and $P = P_{{\rm NM}} + P_{{\rm DM}}$ 
are the local rest-mass density, thermal energy density 
and pressure. $p_{{\nu}}$ is the local neutrino pressure
but we set it to be zero in Eq. \ref{eq:TOV}. $\Gamma$ is the generalized Lorenz factor. 
$m_{{\rm TOV}}$ is the effective enclosed mass defined as
\begin{equation}
m_{{\rm TOV}} = 4 \pi \int^r_0 dr' r'^2 \Gamma \left( \rho + e + E + \frac{F}{\Gamma} \right).
\end{equation}
$E$ and $F$ are the neutrino flux terms which are also set to be zero. 

\subsection{Pre-bounce Weak Process}

To include the electron capture process which is 
essential for triggering the collapse, we applied
the parametrized electron capture scheme first proposed
in \cite{Liebendoerfer2005}. The idea is based on the observation
that the local electron fraction $Y_{{\rm e}}$ is a function of baryon density only.
This is a valid approximation because at high density, 
the Fermi energy is sufficiently high that the electron gas is
extremely degenerate. As a result, the electron capture process is sensitive
to the local baryon density but not the local temperature. 

To implement the scheme, we first find the density dependence of the equilibrium 
electron fraction ${\bar Y}_{{\rm e}} (\rho_{{\rm NM}})$
based on simulations with realistic neutrino
transport. Then we generalize this relation to  
the whole star. In each step, 
we compute the new 
electron capture rate defined by
\begin{equation}
\frac{dY_{\rm e}}{dt} = \frac{\bar{Y}_e(\rho_{{\rm NM}}) - Y_{\rm e}}{\delta t}.
\end{equation}
Here $\delta t$ is the time step. 
The electron capture triggers an entropy change computed by
\begin{equation}
T \frac{d s}{dt} = (\mu_e - \mu_n + \mu_p - E_{{\rm esc}}) \left( \frac{dY_{\rm e}}{dt} \right).
\end{equation}
$E_{{\rm esc}} \approx 10$ MeV is the mean energy directly carried away 
by escaped neutrinos per electron capture reaction. 
For 
$\rho_{{\rm NM}} > \rho_{{\rm crit}}
= 2 \times 10^{12}$ g cm$^{-3}$, 
the neutrinos produced from electron capture are 
assumed to be in equilibrium with the matter and are 
instantly absorbed. Thus, there is no entropy change 
for the variation of $Y_{\rm e}$. 
For 
$\rho_{{\rm NM}} < \rho_{{\rm crit}}$, we check if the chemical potentials
$\mu_n$, $\mu_p$, $\mu_e$ for neutrons, protons and electrons respectively
satisfy $\mu_e - \mu_n + \mu_p - E_{{\rm esc}} > 0$.
If the chemical potential condition is not satisfied, no
entropy change is made. 

The neutrinos are assumed to be an ideal degenerate Fermi 
gas and are in equilibrium in neutrino-opaque zones,
providing a pressure 
\begin{equation}
p_{\nu} = \frac{4 \pi}{3} \left( \frac{k_B T}{hc} \right)^3 (k_B T) F_3 \left( \frac{\mu_{\nu}}{k_B T} \right),
\end{equation}
where $\mu_{\nu} = \mu_e - \mu_n + \mu_p$ is the 
chemical potential for the electron neutrinos and $k_B$ is the Boltzmann constant.
The $F_n(\xi)$ is the 
Fermi-Dirac integral of the $n^{{\rm th}}$-order. 
Analytic approximations are used \citep{Epstein1981}.
We refer the interested readers to \cite{Liebendoerfer2005}
for a detailed discussion of the scheme and its implementation
including different treatments in the neutrino 
opaque and transparent zones. 

\begin{figure}
\centering
\includegraphics*[width=8cm,height=6.4cm]{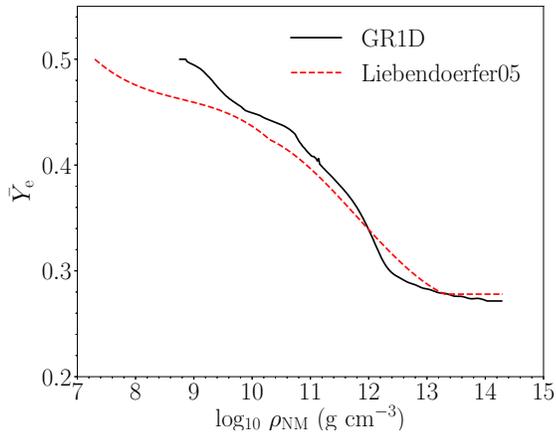}
\caption{The $\bar{Y}_{{\rm e}} (\rho_{{\rm NM}})$ relation using the 
GR1D simulation (this work) and that in \cite{Liebendoerfer2005}
based on detailed neutrino transport.}
\label{fig:ye_rho_plot}
\end{figure}

In this work, we used two parametrizations
to build the $\bar{Y}_{{\rm e}} (\rho_{{\rm NM}})$ relation. 
The first one is the default one, in which we first use
the GR1D code \citep{Connor2010} for a benchmark AIC model based on 
the white dwarf described above but without DM. Then 
we follow the detailed neutrino transport from the 
onset of simulation to trace the evolution of the core electron
fraction as a function of $\rho_c$. We tabulate 
this relation and apply it to other simulations, assuming that 
the matter in the outer region follows the 
same $\bar{Y}_{e}(\rho_{{\rm NM}})$ relation. 
The second method is the parametrized scheme in
\cite{Liebendoerfer2005}. The simulation is based on 
the collapse of a 15 $M_{\odot}$ Fe core using Boltzmann
neutrino transport. 
In Figure \ref{fig:ye_rho_plot} we plot the $\bar{Y}_{{\rm e}} (\rho_{{\rm NM}})$
relations for the two methods we used in our 
simulations. The two
schemes give similar results. The GR1D code gives a slightly
higher $\bar{Y}_{{\rm e}}$ at low density, with a transition at
$\sim 10^{12}$ g cm$^{-3}$.

\section{Results}
\label{sec:results}

In this section we first present the hydrodynamics and 
neutrino transport results for the model without DM. We regard 
as the benchmark model, which 
serves as a verification of the numerical scheme
we have used by comparing with similar models in the literature.

\subsection{Benchmark model}

\begin{figure}
\centering
\includegraphics*[width=8cm,height=6.4cm]{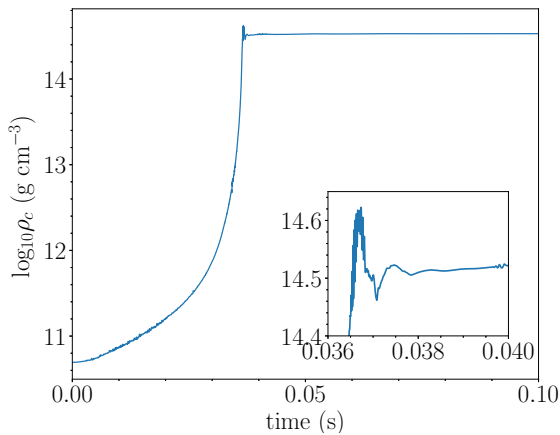}
\caption{Central density against time for the benchmark model.
The smaller panel is the zoomed-in plot around the bounce.}
\label{fig:rhoc_benchmark_plot}
\end{figure}

\begin{figure*}
\centering
\includegraphics*[width=12cm,height=9.6cm]{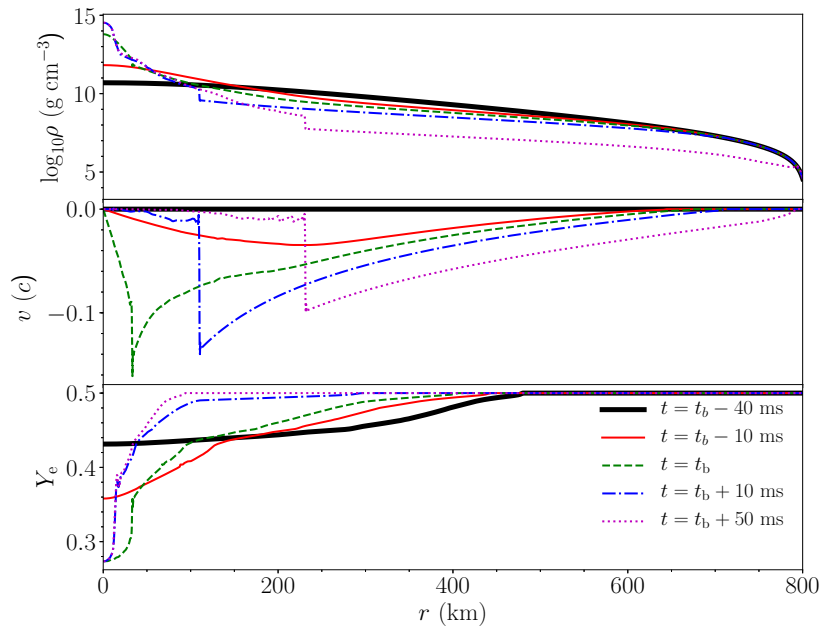}
\caption{The density (upper panel), velocity (middle panel) and $Y_{{\rm e}}$ 
profiles (lower panel) 
for the benchmark model at $t = 40$ ms (black solid line) and 10 ms (red solid line)
before bounce, bounce (green dashed line), 10 ms (blue dot-dashed line) and 
50 ms (purple dotted line) after bounce.}
\label{fig:profiles_benchmark_plot}
\end{figure*}

In Figure \ref{fig:rhoc_benchmark_plot} we plot the central density
against time for the benchmark model, i.e. without DM. 
The electron capture can trigger the 
collapse very quickly. Within 30 ms the central density
increases from its initial value $5 \times 10^{10}$ g cm$^{-3}$
to $\sim 3 \times 10^{14}$ g cm$^{-3}$. After the core
reaches its maximum density, it shows small fluctuations 
in the central density known as the ringdown. Then 
the core density remains roughly constant. 

In the upper panel of Figure \ref{fig:profiles_benchmark_plot} we plot the density 
profiles of the benchmark model at the beginning of the simulation, after 30 ms, 
at bounce, 10 and 50 ms after bounce. The initial profile is that
of a white dwarf 
at hydrostatic equilibrium with uniform 
temperature of 0.01 MeV and $Y_{{\rm e}} = 0.5$. Once the initial electron capture 
takes place $(t = 0)$, the innermost 300 km, where $Y_{{\rm e}}$ is lowered,
starts to contract. The outer material also contracts but in a slower
manner. 
At the bounce, which is defined by the moment when the entropy
per baryon at the edge of the inner core exceeds 3 $k_B$, 
the PNS with an envelope appears. 
The innermost 100 km is the PNS
where the density is above $10^{11}$ g cm$^{-3}$. 
In the outer part matter fall inwards adiabatically. 
At 10 ms after bounce, a bounce shock clearly
developed at $\sim 100$ km. The PNS density does
not change significantly.

In the middle panel of Figure \ref{fig:profiles_benchmark_plot} we plot the velocity profiles. 
The initial velocity profile
is everywhere zero by construction. At 30 ms after the first electron
capture, the infalling velocity of matter is increasing owing
to the suppression of pressure by electron capture. Matter at 200 km
from the core has the highest infalling velocity at 0.05 $c$. 
At the bounce, a sharp velocity cusp is observed at $\sim 20$ km,
which is the boundary of the PNS core where the matter reaches
nuclear matter density. The bounce shock can be seen to propagate 
by comparing the profiles between bounce and 10 ms after the bounce.  
The velocity at the bounce shock can be as high as 0.15 $c$ falling inwards, while
the matter interior of the bounce shock is close to being static.

The lower panel of Figure \ref{fig:profiles_benchmark_plot} is for the 
$Y_{\rm e}$ profiles. The initial $Y_{{\rm e}}$ is lowered
at the WD center with the minimum at 0.43. Before the bounce, 
$Y_{{\rm e}}$ has the largest reduction at the core
and can be as low as 0.35. During the
bounce, $Y_{{\rm e}}$ reaches its minimum of 0.27.
At 10 ms after bounce, the $Y_{{\rm e}}$ profile only 
moves inward. 
Since we do not have neutrino transport
in our modeling, the parametrized electron capture is
turned off after bounce when the exact $Y_{{\rm e}}$ profile
depends strongly on the competition
between neutrino emission and absorption.
The $Y_{{\rm e}}$ outside
the PNS can also increase by neutrino absorption. 
As a result, the $Y_{{\rm e}}$ profile after bounce is 
only for reference. 

\subsection{Effects of Dark Matter Admixture}

After discussing the benchmark model, we study how the admixture of
DM affects the collapse and bounce process and the resultant NS. 
In Table \ref{table:models} we tabulate the initial configurations,
global properties of the collapse dynamics and the PNS after the bounce. 

\begin{table*}
\caption{Models used in this work. $\rho_{c{\rm (NM),ini}}$
and $\rho_{c{\rm (DM),ini}}$ are the central densities of baryonic matter
and DM in units of $10^{10}$ g cm$^{-3}$. $\rho_{{\rm b}}$ is
the maximum baryonic density reached
in unit of $10^{14}$ g cm$^{-3}$. $R_{{\rm NM}}$ and $R_{{\rm DM}}$ are
the initial radii of the baryonic matter and DM in unit
of km. $T_{{\rm ini}}$ is the initial central temperature in unit of MeV. 
$t_{{\rm b}}$ is the time when the core reaches a
central density of $10^{14}$ g cm$^{-3}$ in unit of ms. $M$, $M_{{\rm NM}}$, $M_{{\rm DM}}$ and $M_{{\rm NS}}$ 
are the masses of the initial model, its baryon and DM components and 
the proto-neutron star in unit of $M_{\odot}$,
defined as the enclosed mass with a central density above $10^{11}$ g cm$^{-3}$.
"nil" is for the cases that the models fail to collapse into a neutron star.}
\begin{center}
\begin{tabular}{|c|c|c|c|c|c|c|c|c|c|c|c|c|c|}
\hline
Model & $\rho_{c{\rm (NM),ini}}$ & $\rho_{c{\rm (DM),ini}}$ & $T_{{\rm ini}}$ & $M$ & $M_{{\rm NM}}$ & $M_{{\rm DM}}$ & $R_{{\rm NM}}$ & $R_{{\rm DM}}$ & 
$t_{{\rm b}}$ & $\rho_{{\rm b}}$ & EOS & Remarks  \\
\hline
5-0-c-SFHo-G & 5 & 0 & 0.01 & 1.35 & 1.35 & 0.00 & 810 & 0 & 40 & 3.4 & SFHo & \\
5-1-c-SFHo-G & 5 & 26.1 & 0.01 & 1.31 & 1.30 & 0.01 & 880 & 50 & 43 & 3.2 & SFHo & \\
5-2-c-SFHo-G & 5 & 50.0 & 0.01 & 1.28 & 1.26 & 0.02 & 980 & 50 & 53 & 3.1 & SFHo & \\
5-3-c-SFHo-G & 5 & 91.9 & 0.01 & 1.24 & 1.21 & 0.03 & 1120 & 40 & 73 & 2.9 & SFHo & \\
5-4-c-SFHo-G & 5 & 148 & 0.01 & 1.19 & 1.15 & 0.04 & 1340 & 40 & 119 & 2.8 & SFHo & \\
5-5-c-SFHo-G & 5 & 222 & 0.01 & 1.13 & 1.08 & 0.05 & 1710 & 40 & 306 & 2.6 & SFHo & \\
5-6-c-SFHo-G & 5 & 313 & 0.01 & 1.05 & 0.99 & 0.06 & 2400 & 40 & nil & nil & SFHo & \\ \hline 

5-0-c-LS220-G & 5 & 0 & 0.1 & 1.40 & 1.40 & 0.00 & 830 & 0 & 40 & 4.1 & LS220 & \\
5-2-c-LS220-G & 5 & 50.0 & 0.1 & 1.34 & 1.32 & 0.02 & 1010 & 50 & 53 & 3.5 & LS220 & \\
5-4-c-LS220-G & 5 & 148 & 0.1 & 1.23 & 1.19 & 0.04 & 1640 & 40 & 139 & 3.2 & LS220 & \\
5-5-c-LS220-G & 5 & 219 & 0.1 & 1.18 & 1.13 & 0.05 & 2200 & 40 & 310 & 3.0 & LS220 & \\
5-6-c-LS220-G & 5 & 313 & 0.1 & 1.15 & 1.06 & 0.06 & 3580 & 40 & nil & nil & LS220 & \\ \hline 

5-0-c-HShen-G & 5 & 0 & 0.01 & 1.35 & 1.35 & 0.00 & 810 & 0 & 35 & 2.6 & HShen & \\ \hline

5-0-h-SFHo-G & 5 & 0 & 0.86 & 1.42 & 1.42 & 0.00 & 810 & 0 & 36 & 3.1 & SFHo & \\ \hline

5-0-c-SFHo-L & 5 & 0 & 0.01 & 1.35 & 1.35 & 0.00 & 810 & 0 & 33 & 3.0 & SFHo & \cite{Liebendoerfer2005} \\
5-0-c-SFHo-G-Newt & 5 & 0 & 0.01 & 1.36 & 1.36 & 0.00 & 810 & 0 & 33 & 2.9 & SFHo & Newtonian gravity \\ \hline
5-0-c-SFHo-G-coarse & 5 & 0 & 0.01 & 1.35 & 1.35 & 0.00 & 810 & 0 & 36 & 3.9 & SFHo & $\Delta r = 0.8$ km\\ 
5-0-c-SFHo-G-fine & 5 & 0 & 0.01 & 1.35 & 1.35 & 0.00 & 810 & 0 & 36 & 3.7 & SFHo & $\Delta r = 0.2$ km\\ \hline
\end{tabular}
\label{table:models}
\end{center}
\end{table*}

We use model names that contain the necessary model parameters.  
For example, for Model 5-6-c-SFHo-G, the first
entry (5) corresponds to the initial central density of baryonic 
matter in unit of $10^{10}$ g cm$^{-3}$. The second entry (6) is the 
DM total mass in unit of 0.01 $M_{\odot}$. The third entry (c) 
indicates whether it is cold c or hot h. 
The fourth entry (SFHo) reveals the EOS, SFHo, LS220 or HShen. The last entry (G) is the electron capture
parametrization, where G is the scheme by the GR1D
code and L is that in \cite{Liebendoerfer2005}.

\begin{figure}
\centering
\includegraphics*[width=8cm,height=6.4cm]{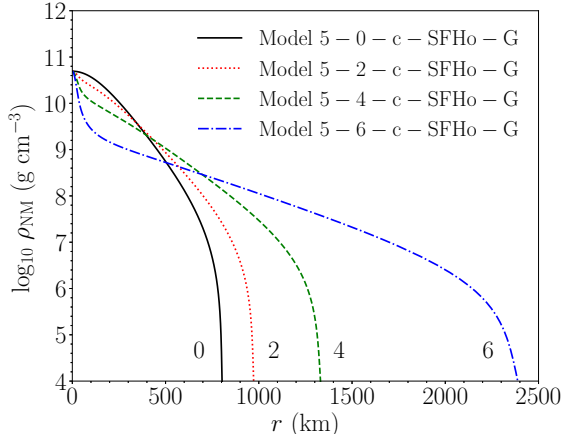}
\caption{Initial density profiles of Models 5-0-c-SFHo-G (black solid line), 
5-2-c-SFHo-G (red solid line), 5-4-c-SFHo-G (green dashed line) and 5-6-c-SFHo-G
(blue dot-dashed line)}
\label{fig:rho_ini_DM_plot}
\end{figure}

In Figure \ref{fig:rho_ini_DM_plot} we plot the initial profiles
for some of the models. We refer the readers to our previous
work \citep{Leung2013} for more detailed profiles of WD with 
admixed DM in hydrostatic equilibrium. It can be seen that,
as remarked in Table \ref{table:models}, the radius of a WD
is larger for a more massive admixed DM. A two-layer structure 
can be seen. For the innermost layer ($\sim$ 40 km), where the DM locates, 
the baryon density drops rapidly. Outside that the density profile falls
slower than models with less admixed DM because the gravity is dominated
by the baryons.
The radii of the stars with and without DM admixed can 
differ by three times, as seen by comparing Models 5-0-c-SFHo-G and 5-6-c-SFHo-G.

\begin{figure}
\centering
\includegraphics*[width=8cm,height=6.4cm]{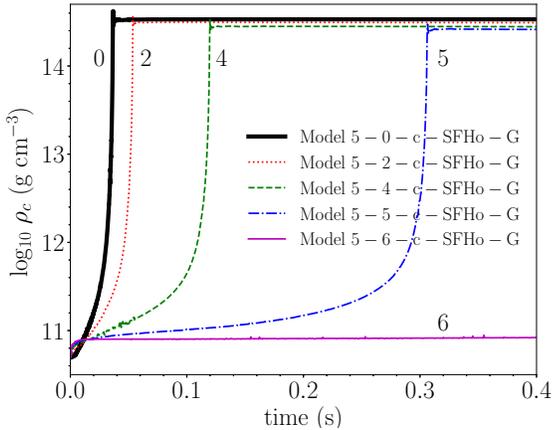}
\caption{Central baryon densities of Models 5-0-c-SFHo-G (black solid line), 
5-2-c-SFHo-G (red dotted line),
5-4-c-SFHo-G (green dashed line), 5-5-c-SFHo-G (blue dot-dashed line)
and 5-6-c-SFHo-G (purple solid line) respectively.}
\label{fig:rho_DM_plot}
\end{figure}

\begin{figure}
\centering
\includegraphics*[width=8cm,height=6.4cm]{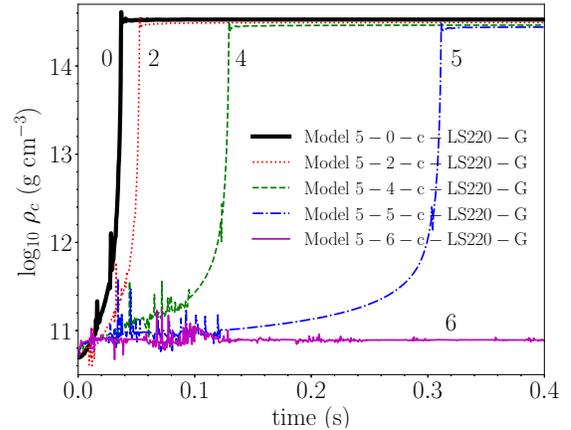}
\caption{Central baryon densities of Models 5-0-c-LS220-G (black solid line), 5-2-c-LS220-G (red dotted line),
5-4-c-LS220-G (green dashed line), 
5-5-c-LS220-G (blue dot-dashed line) and 5-6-c-LS220-G (purple solid line) respectively.}
\label{fig:rho_DM_plot2}
\end{figure}

We plot in Figure \ref{fig:rho_DM_plot} the central baryon density
against time for Models 5-0-c-SFHo-G, 5-2-c-SFHo-G,
5-4-c-SFHo-G, 5-5-c-SFHo-G and 5-6-c-SFHo-G. These models 
differ from each other by the mass of the admixed DM. 
When the initial WD has more admixed DM, even when initial central densities 
of baryonic
matter are the same, the time needed
for the core to reach the nuclear density and bounce
increases. It can grow from the standard case $\sim 40$ ms
to as long as $\sim 300$ ms when $M_{{\rm DM}} = 0.05 ~M_{\odot}$. 
For $M_{{\rm DM}} = 0.06$ $M_{\odot}$, the central density
increases for $\sim 1$ s and then gradually drops.
This signifies that the core fails to collapse. 
This is because the gravity of the admixed DM becomes large enough 
that the density of baryonic matter drops significantly 
fast in the inner core
region (see Figure \ref{fig:rho_ini_DM_plot}).
As a result, the region for efficient electron capture becomes 
too small for triggering a global collapse.

To contrast with the softer SFHo EOS, in Figure \ref{fig:rho_DM_plot2} we also 
plot the central density evolution of Models 5-0-c-LS220-G,
5-2-c-LS220-G, 5-4-c-LS220-G, 5-5-c-LS220-G and 5-6-c-LS220-G
which are models differing only by the 
admixed DM masses, with all models using the LS220 EOS. 
Models with $M_{{\rm DM}} < 0.06$ can collapse
into a NS. The bounce time is delayed when $M_{{\rm DM}}$ 
increases. There are more spikes in the central
density evolution, showing that unlike the SFHo counterpart,
the star generates more sound waves during its collapse.
However, after bounce, the PNS quickly reaches its equilibrium
and the central density reaches an equilibrium value,
which decreases when $M_{{\rm DM}}$ increases. 

\begin{figure}
\centering
\includegraphics*[width=8cm,height=6.4cm]{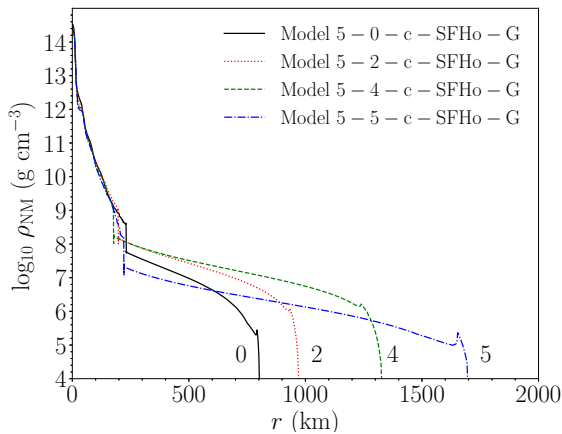}
\caption{Baryon density profiles of Models 5-0-c-SFHo-G (black solid line), 
5-2-c-SFHo-G (red dotted line),
5-4-c-SFHo-G (green dashed line) and 5-5-c-SFHo-G (blue dot-dashed line) respectively at 50 ms after the bounce. The numbers next to the lines label the amount of DM admixed (in 0.01 $M_{\odot}$)}
\label{fig:rho_DM_bounce_plot}
\end{figure}

To further extract the effects of the DM admixture, 
we plot in Figure \ref{fig:rho_DM_bounce_plot} the density
profiles at 50 ms after bounce. Even though detailed neutrino transport
is needed for an accurate description of the full evolution, the early development
after bounce should be valid because the bounce shock is in 
the region which is opaque to most neutrinos, so that they are in equilibrium
with the matter. From the figure, we can see that after
bounce the density structures for the four models are very similar.
A bounce shock locates at $\sim 200$ km from the core. 
The four models
have a similar envelope structure, the density gradient of which
decreases with increasing $M_{{\rm DM}}$. 

\begin{figure}
\centering
\includegraphics*[width=8cm,height=5.4cm]{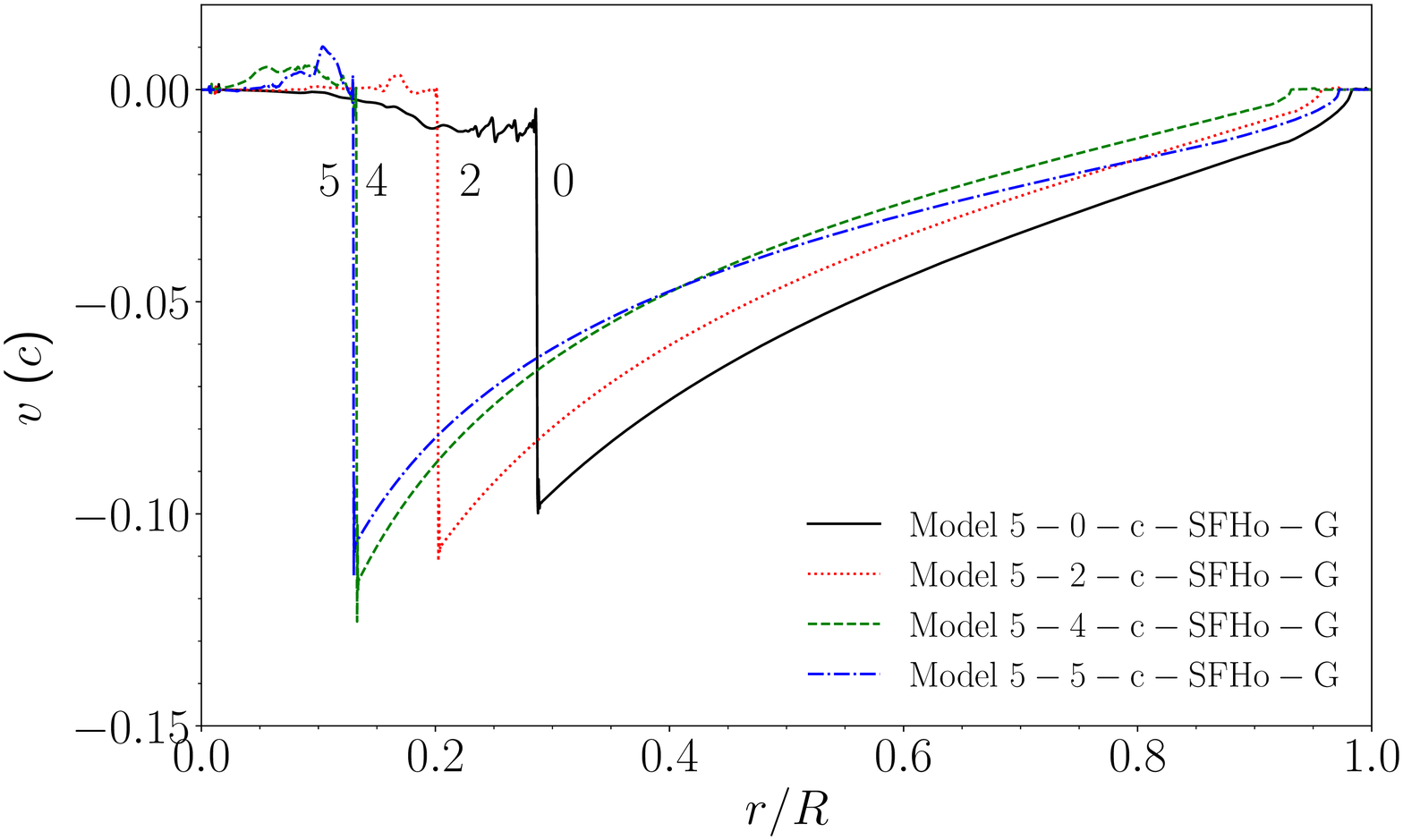}
\caption{Same as Figure \ref{fig:rho_DM_bounce_plot}, but for the velocity profiles
and the $r$-coordinate scaled with the corresponding stellar radius.}
\label{fig:vel_DM_bounce_plot}
\end{figure}

We also plot in Figure \ref{fig:vel_DM_bounce_plot} the velocity
profiles at 50 ms after bounce for the same set of models as in
Figure \ref{fig:rho_DM_bounce_plot}. The velocity profiles also 
behave very similarly among the models. This suggests that the dynamics
after bounce is dominated by the matter with nuclear density, which is 
less sensitive to the pre-collapse structure and the total mass of the 
initial WD $M$. 

\begin{figure}
\centering
\includegraphics*[width=8cm,height=6.4cm]{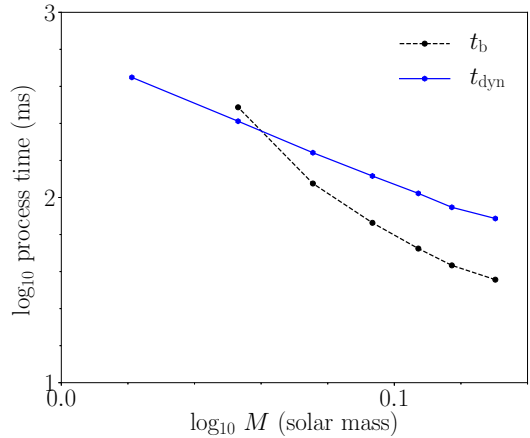}
\caption{Bounce time against $M$ (dashed line) based
on Models 5-x-c-SFHo-G
where $x = 0, 1, 2, 3, 4$ and 5 and the corresponding
dynamical time calculated with the initial radius and mass (solid line).}
\label{fig:collapse_time}
\end{figure}

At last we plot $t_b$ and the dynamical time against $M$ 
in Figure \ref{fig:collapse_time} for the models presented
in this work. The dynamical time $t_{{\rm dyn}} = \sqrt{2 R^3 / G M}$ is defined by the 
initial mass and radius of the WD.
As discussed in the previous section, the 
$t_b$ increases in a quasi-exponential manner 
from $\sim 30$ ms for $M_{\rm DM}$ = 0 up to
$\sim 300$ ms for $M_{\rm DM}$ = 0.05 $M_{\odot}$.

For more massive admixed DM (e.g., Model 5-6-c-SFHo-G), 
WD fails to collapse into a NS. 
The central density of the 
WD increases and then decreases later at $\sim$1 s for this model,
where the central density never exceeds
the neutron drip density. This suggests that the
admixed DM is sufficiently massive to stop further collapse,
even when we impose the initial $Y_{{\rm e}}$ change.
It is because the admixed DM core creates a steep density profile 
in the initial data (see Figure \ref{fig:rho_ini_DM_plot}). Only a 
small core can carry out electron capture in the first place.
On the other hand, for those models
that lead to NS formation, we observe
a very steep correlation between 
the progenitor mass and the collapse time, 
\begin{equation}
t_b = 1090 M^{-12},
\end{equation}
where $t_b$ is in unit of ms and $M$ is in unit of
$M_{\odot}$.

\subsection{Dependence of AIC on Input Physics}

\begin{figure}
\centering
\includegraphics*[width=8cm,height=6.4cm]{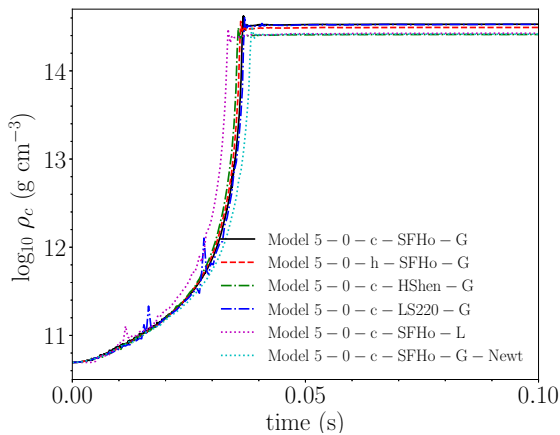}
\caption{Central baryon densities of Models 5-0-c-SFHo-G (black solid line), 
5-0-h-SFHo-G (red dashed line), 
5-0-c-Hshen-G (green dot-dashed line), 5-0-c-LS220-G (blue dot dashed line),
5-0-c-SFHo-L (purple dotted line) and 5-0-c-sFHo-G-Newt (cyan
dotted line) vs. time.}
\label{fig:rhoc_comp_plot}
\end{figure}

In this section,
we examine how several input physics parameterizations affect the collapse
dynamics and properties of the resulting PNS.

\subsubsection{Initial temperature}

It is known that for AIC, especially for C-O WD's, the pre-collapse
dynamics can be complicated because of the nuclear deflagration,
which releases thermal energy to heat up and thermalize the core before 
collapse. To mimic the nuclear burning before 
collapse, we compare in Figure \ref{fig:rhoc_comp_plot} Models
5-0-c-SFHo-G and 5-0-h-SFHo-G,
which have the same configuration except for the temperature profile.
The $c$-model has an isothermal temperature 
at 0.01 MeV, while the $h$-model has a density dependent 
temperature profile with a central temperature of 0.86 MeV. The chosen value for the central
temperature corresponds when the matter burns
into nuclear statistical equilibrium.

In Figure \ref{fig:rhoc_comp_plot} we plot the central baryon density
against time for the two models. Despite the 
initial masses differing by $\sim 8 \%$, the collapse dynamics
are almost the same because of the strong degeneracy of matter. 
The $h$-model has a slightly lower central density and faster
collapse time.

\subsubsection{Equation of State}

The nuclear matter
EOS is highly uncertain due to the lack of direct 
constraints on neutron star radius and compressibility. 
Despite that, a number of EOSs,
such as the SFHo EOS \citep{Steiner2013}
can produce NS with mass as high as that of the recently 
observed pulsar PSR J0348+0432. To understand
the effect of nuclear matter EOS on the bounce dynamics,
we pick two other EOSs, the HShen EOS and LS220 EOS
to compare with. We choose these EOSs because they are 
used extensively, such as in the AIC simulations in \cite{Abdikamalov2010}.

WD using the LS220 EOS in general has a higher 
mass. The radius of the WD does not differ much for the 
same $M$ and $M_{{\rm DM}}$ between the LS220 and SFHo EOSs.
The WD models using the LS220 EOS have a slightly higher 
$t_b$ when $M_{{\rm DM}}$ is high, otherwise almost 
no difference from the SFHO EOS. At last, the 
maximum density reached during bounce is higher 
for LS220 than SFHo EOS. Despite these differences,
the qualitative trends for the initial $M$, 
$t_b$ and $\rho_b$ against $M_{{\rm DM}}$ are identical for the two EOSs.

We plot in Figure \ref{fig:rhoc_comp_plot} the central 
density against time for Models 5-0-c-SFHo-G, 5-0-c-LS220 and 5-0-c-HShen-G. 
The three models are again identical in most 
parts except for the choice of EOS. We can see that 
the collapse of the HShen-model is faster with a lower central density for the
PNS. The LS220-model collapses slightly slower than other two models.
However, the difference is very small
($\sim 10 \%$). Therefore, we conclude that 
the nuclear EOS plays a less important
role in the bounce dynamics and the PNS in the AIC scenario.

\subsubsection{Electron Capture Scheme}

We have chosen to use our own $\bar{Y}_{{\rm e}} (\rho_{{\rm NM}})$
relation for the parametrized electron capture, because
the electron capture in AIC can be different from that in 
core-collapse supernovae (CCSNe). 
As remarked in Figure \ref{fig:ye_rho_plot} the one from CCSNe has a 
lower (higher) $Y_{{\rm e}}$ at lower (higher) density than the AIC model. 

We plot in Figure \ref{fig:rhoc_comp_plot} the central density against
time for Models 5-0-c-SFHo-G and 5-0-c-SFHo-L. The two models 
are similar except that the parametrized electron capture in the latter model 
is taken directly from \cite{Liebendoerfer2005}. We can see that 
the $L$-model has a faster 
collapse time and lower PNS central density by $\sim 10 \%$. 
However, the collapse dynamics remain similar despite the 
different electron capture schemes used.

\subsubsection{Newtonian Gravity}

We examine the effect of the gravity solver on the collapse
dynamics. It is natural to use the general relativistic (GR) gravity solver
to model the collapse involving the presence of a NS. However,
in general the relativistic effect of the NS is still small,
as $G M_{{\rm NS}} / c^2 R_{{\rm NS}} \sim 0.1$. 
Here we take $M_{{\rm NS}} \approx 1.4 ~M_{\odot}$ and
$R_{{\rm NS}} = 15$ km.  
It therefore becomes interesting to see how large 
a difference in the collapse dynamics the 
GR corrections make.

In Figure \ref{fig:rhoc_comp_plot} we plot the central 
density against time for Models 5-0-c-SFHo-G and 5-0-c-SFHo-G-Newt, 
corresponding to the models using the  
approximate GR gravity solver and the Newtonian 
gravity solver \citep{Leung2015a} respectively. We can see that 
the Newtonian one shows a slower collapse and lower
final central density. This is consistent with the fact
that GR includes the pressure in the gravity source term
which deepens the gravitational well compared to Newtonian gravity. 
Despite that the changes in the bounce time and 
final central baryon density are only about 10 \%.

\section{Discussion}
\label{sec:discussion}

\subsection{Connection to Observed Neutron Star Mass}

We have examined the collapse process of AIC with admixed
DM. We show that AIC models with a mass above 1.11 $M_{\odot}$
collapse into NSs. In this section, we 
attempt to connect the observed NS mass with the admixed 
DM. For AIC models, it has been shown that, due to the lack of 
an extended envelope, the matter outside the PNS
can barely escape. As reported in \cite{Metzger2009a},
the escape mass for a typical AIC event is as low as 
$\sim 10^{-2} ~M_{\odot}$. Therefore, 
the PNS mass is a good approximation of the final NS mass. 
The gravitational mass of the NS can be further 
lowered compared to the baryonic mass by another $10 \%$. 

\begin{figure}
\centering
\includegraphics*[width=8cm,height=6.4cm]{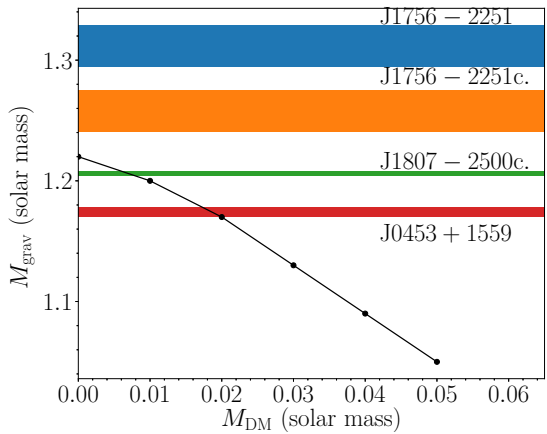}
\caption{$M_{{\rm grav}}$ against $M_{{\rm DM}}$ 
of this work for the SFHo models. Selected pulsar mass data are presented,
including J1756-2251 and J1756.2251c. \citep{Ferdman2014}, 
J1807-2500c. \citep{Lynch2012} and the 
J0453+1559 \citep{Martinez2015}.}
\label{fig:ns_mass_plot}
\end{figure}
 
In Figure \ref{fig:ns_mass_plot} we plot the progenitor gravitational 
mass $M_{{\rm grav}}$ against $M_{{\rm DM}}$, together with a few
well observed pulsars with a small error bar in the mass. We follow \cite{Prakash2001,Suwa2018} 
to estimate $M_{{\rm grav}}$ from the total mass by solving
\begin{equation}
M - M_{{\rm grav}} - 0.084 M_{\rm grav}^2 = 0,
\end{equation}
with all masses in units of $M_{\odot}$.
We also note that this only gives a preliminary estimate because
the formula is originally designed for NS with pure baryons
and the coefficient 0.084 is obtained from a specific 
EOS. 
The choices of pulsars include J1756-2251 and its companion \citep{Ferdman2014}, 
companion of J1807-2500 \citep{Lynch2012} and 
J0453+1559 \citep{Martinez2015}. The 
$M_{{\rm grav}}$ range predicted in this work matches well for 
pulsars with mass below 1.22 $M_{\odot}$. 
The neutron star with the lowest mass so far
J0453+1559 requires at most 0.02 $M_{\odot}$ admixed DM 
in the WD prior to its collapse. 
More massive pulsars including the pulsar binary J1756-2251
are probably produced from traditional core-collapse supernovae.

It should be noted that the formation of a
low-mass NS is difficult from the stellar evolution perspective. 
An MS star with mass above 10 $M_{\odot}$ has a Fe core exceeding the Chandrasekhar 
mass $\sim 1.4 ~M_{\odot}$. These stars are known to have 
a smooth density gradient between the 
Fe-core and the outer envelope. 
Hence, even if the core collapse may result in an
explosion as a core-collapse supernova, the accretion of matter from
the outer envelope and the later fallback of the inner
material will only increase the final mass of the PNS. 
As a result, the PNS's produced from these MS stars
cannot have masses matching those of the observed low-mass NS's.

For an MS star with mass 8 - 10 $M_{\odot}$, 
an ONeMg core forms instead of a Fe-core and it
does not exceed the Chandrasekhar mass. 
The core can have a central density as high as $\sim 10^{10}$ g cm$^{-3}$,
but the H-envelope is detached from the core as characterized by
the steep density gradient. 
The electron captures of $^{24}$Mg$(e^-,\nu_e)^{24}$Na$(e^-,\nu_e)^{24}$Ne 
and $^{20}$Ne$(e^-,\nu_e)^{20}$F$(e^-,\nu_e)^{20}$O 
trigger the collapse of the ONeMg core \citep{Nomoto2017a,Leung2018b}.
In this scenario, the final PNS mass can be lower than the 
core-collapse case. However, from multi-dimensional simulations
with neutrino transport (see, e.g. \cite{Dessart2006}), the mass
ejection in these events is very small ($\sim 10^{-2-3} ~M_{\odot}$).
This means that the final PNS mass clusters near the Chandrasekhar mass of the 
ONeMg core ($\sim 1.3 ~M_{\odot}$). Recent one-dimensional
survey done in \cite{Sukhbold2016} shows that the final masses of the stellar 
remnants using the Z9.6 engine are from
1.35 - 1.53 $M_{\odot}$. Recent two- (three-) dimensional simulations \citep{Burrows2019}
of 9 and 10 $M_{\odot}$ stars also give similar results 
of 1.358 (1.342) and 1.524 (1.495) $M_{\odot}$ respectively.
These simulations demonstrate that the minimal baryonic mass for 
the remnant NS is $\sim 1.36 ~M_{\odot}$ in
the single star scenario. This corresponds to the 
gravitational mass of 1.24 $M_{\odot}$. 

In \cite{Suwa2018}, 
the possibility that a low-mass Fe-core evolved from stellar models 
of mass from 8.8 to 9.3 $M_{\odot}$ is considered
(corresponding to CO core mass from 1.37 to 1.45 $M_{\odot}$). 
The resultant gravitational mass of the NS ranges from
1.17 to 1.25 $M_{\odot}$. From their hydrodynamics simulations, 
they further remarked that the collapse scenario in general
results in very low mass loss. The pulsar J1756-2251 can therefore
be explained by their models only marginally. 

In both channels for PNS formation by the standard 
stellar evolutionary paths, the PNS mass is either higher than the standard
Fe-core Chandrasekhar mass or around the ONeMg core Chandrasekhar mass.
Therefore, it becomes challenging in the stellar evolution point of view
to form the observed low-mass pulsar as described above. On the other hand,
DM admixture provides 
a robust mechanism for formation of the 
initial sub-Chandrasekhar mass core, which can undergo AIC to generate a low-mass PNS.

\subsection{Effects of Dark Matter Particle Mass}

In the above, we have only studied the
effects of 1 GeV DM particles as motivated by
e.g., \cite{Foot1991, Okun2007}. 
However, depending on the DM model, the 
DM particle mass can be very different from 1 GeV.
Here we briefly discuss the 
effects of particle mass on the collapse scenario. 

In general, the Chandrasekhar mass of the 
admixed DM decreases when the DM particle mass $m_{{\rm DM}}$ increases
due to the $1/m_{{\rm DM}}$ dependence of the Fermi pressure.
As reported in \cite{Leung2013},
DM with a high particle mass will have a smaller effect
on the Chandrasekhar mass of the admixed WD. Therefore, the corresponding
NS mass range will be smaller and closer to the canonical value.
The corresponding 
Chandrasekhar mass of the DM counterpart, which scales
as $1 / m_{{\rm DM}}^2$, may be more easily exceeded.
If this is satisfied, the collapse of the DM core may trigger 
the further collapse into a BH,
and disrupts the host star.
The deposited energy by radiation
may also trigger the explosion of the WD \citep{Graham2015}.
The NS is likely to be disrupted directly \citep{Goldman1989}.
However, dynamical simulation of axion DM suggests
that accretion exceeding the Chandrasekhar mass of the DM 
core is less likely due to the larger mass loss during 
accretion \citep{Brito2015}. If the DM core
can reach such a limit, this is one of the possible channels of triggering the 
fast radio burst discussed in \cite{Fuller2015}.

On the other hand, the DM component with a sufficiently small DM particle mass
has a size larger
than that of the baryonic counterpart. Such an object will go through
very unusual collapse. The gravitational well
can be dominated by the DM instead of baronyic matter \citep{Leung2011}.
The collapse process is dominated by the baryonic physics by electron capture. However, 
the ejecta (e.g. $^{56}$Ni) and its corresponding light curves,
will be completely different because the ejecta
can no longer escape freely. The ejecta is dominated by the extended and deepened 
gravitational well due to the DM. Depending on its
velocity, the ejecta can still be trapped.

\subsection{Conclusion and Future Works}

In this article we have studied the effects
of DM admixture in the AIC of a WD by using one-dimensional hydrodynamics
simulations. In \cite{Leung2015b} we have considered 
Type Ia supernovae as the complementary part of the picture.
For stars with a larger progenitor
mass and with a larger binding energy (O-Ne-Mg WD compared
with C-O WD), the mass accretion from a companion star
is more likely to trigger a cold collapse of the white dwarf. 

In this work, the DM is modeled as a
non-self-annihilating ideal degenerate 
Fermi gas with a particle mass of 1 GeV
in the simulations. 
We construct the WD in hydrostatic 
equilibrium for both normal matter and DM
with the admixed DM mass up to 0.06 $M_{\odot}$. 
Then we follow its collapse until the formation
of a PNS after a few tenths of ms. 

The admixed DM tends to slow down the collapse
event which results in a lower mass neutron star. 
We show that the lowered NS mass
is a robust result, with respect to the initial temperature, nuclear matter EOS, electron capture scheme, and general relativistic effects, as long as the WD contains the 
admixed DM. 
The mass range of the NS produced in our scenario is compatible with those of  
observed NSs with sub-Chandrasekhar mass. 
These low-mass pulsars can be originated from a WDs
with admixed DM of $\sim 0.01 ~M_{\odot}$. 

In this work we have not considered the roles of neutrinos
in the collapse. It will be interesting to follow 
the evolution of the collapse and bounce with a suitable neutrino transport to see
how the bounce shock propagates, especially in the 
regime where the density profile is very different
from the one without DM. This can lead
to observable differences. Furthermore, the extension of DM to different
particle masses, especially to sub-GeV scale, will also give 
rise to very unusual collapse events.

\section{Acknowledgment}

This work was supported by World Premier 
International Research Center Initiative 
(WPI), MEXT, Japan and JSPS KAKENHI Grant Numbers 
JP26400222, JP16H02168, JP17K05382, and a grant
from the Research Grant Council of Hong Kong (Project
14300317).
We acknowledge the support by the Endowed Research
Unit (Dark Side of the Universe) by Hamamatsu Photonics K.K.
We thank C. Ott and E. O'Connor
for the EOS driver for reading the HShen EOS and
SFHo EOS and also their open source data 
for the parametrized electron capture. 

\appendix

\section{Test of the simulation resolution}

The resolution is known to be a difficult problem in
hydrodynamics simulations,
especially in multi-dimensional ones. It is because halving the mesh size
will result in $2^{N+1}$ times longer running time,
where $N$ is the number of spatial dimensions in the simulation. This poses 
a very tedious constraint on the possible mesh size, 
especially when there are sub-grid physics adopted in the simulation. 
To make sure our simulation does not depend sensitively on
resolution (the default is 0.25 code unit $\approx$ 0.4 km), 
we perform a comparison test for the same models but with 
different mesh sizes. 

\begin{figure}
\centering
\includegraphics*[width=8cm,height=6.4cm]{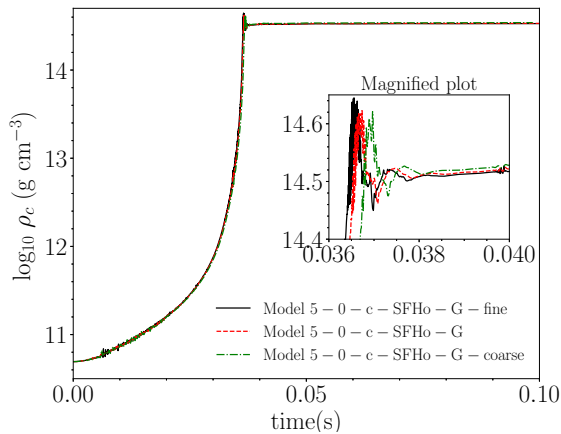}
\caption{Central densities of Models 5-0-c-SFHo-G-coarse, 5-0-c-SFHo-c and
5-0-c-SFHo-G-fine respectively.}
\label{fig:rhoc_dx_plot}
\end{figure}

In Figure \ref{fig:rhoc_dx_plot} we plot the central density
against time for Models 5-0-c-SFHo-G-coarse, 5-0-c-SFHo-c and
5-0-c-SFHo-G-fine respectively. They correspond to 
mesh sizes of $\sim 0.2$, 0.4 and 0.8 km respectively. 
We see that the results agree very well.
There is no significant difference in the central density and the bounce time. 
The differences converge when the mesh size is reduced.
This shows that our current resolution is sufficient
to model the AIC until the formation of the PNS. 
In the low resolution run, the code 
fails to capture the shock at 50 ms after bounce. 
Therefore the minimum resolution to follow the bounce shock
until it stalls is $\sim 0.4$ km. 

\bibliographystyle{apj}
\pagestyle{plain}
\bibliography{biblio}

\end{document}